\documentclass[english,two column, show pacs]{revtex4}
\usepackage[T1]{fontenc}
\usepackage[latin9]{inputenc}
\usepackage{amssymb}
\usepackage{graphicx}
\usepackage{esint}

\makeatletter
\@ifundefined{textcolor}{}
{%
 \definecolor{BLACK}{gray}{0}
 \definecolor{WHITE}{gray}{1}
 \definecolor{RED}{rgb}{1,0,0}
 \definecolor{GREEN}{rgb}{0,1,0}
 \definecolor{BLUE}{rgb}{0,0,1}
 \definecolor{CYAN}{cmyk}{1,0,0,0}
 \definecolor{MAGENTA}{cmyk}{0,1,0,0}
 \definecolor{YELLOW}{cmyk}{0,0,1,0}
}

\makeatother

\usepackage{babel}
\begin{document}

\title{Entanglement of spin-orbit qubits induced by Coulomb interaction}

\author{Y. N. Fang$^{1}$, Yusuf Turek$^{1}$, J. Q. You$^{2}$, and C. P.
Sun$^{1,2}$}

\email{cpsun@csrc.ac.cn}

\affiliation{$^{1}$State Key Laboratory of Theoretical Physics, Institute of
Theoretical Physics, Chinese Academy of Sciences, and University of
the Chinese Academy of Sciences, Beijing 100190, China\\
 $^{2}$Beijing Computational Science Research Center, Beijing 100084,
China}
\begin{abstract}
Spin-orbit qubit (SOQ) is the dressed spin by the orbital degree of
freedom through a strong spin-orbit coupling. We show that Coulomb
interaction between two electrons in quantum dots located separately
in two nanowires can efficiently induce quantum entanglement between
two SOQs. The physical mechanism to achieve such quantum entanglement
is based on the feasibility of the SOQ responding to the external
electric field via an intrinsic electric dipole spin resonance.
\end{abstract}


\maketitle

\section{INTRODUCTION}

Entanglement of particles now has been considered as quantum resource
to implement protocols of quantum information processing \cite{Entangle 1}.
To achieve quantum entanglement for various kinds of qubits is a central
task in quantum information science and technology. With very strong
spin-orbit coupling (SOC), an electron spin becomes the so called
spin-orbit qubit (SOQ) \cite{SOQ 1}, which has been experimentally
implemented in quantum nanowires. Its bit states essentially are two
dressed spin states incorporating with the orbital degree of freedom
(DOF), and thus could feasibly respond to both electric and magnetic
fields. In this sense quantum manipulations on SOQ can be achieved
via electron-dipole spin resonance \cite{Rui Li,EDSR}. 

With those existing progresses, we will face with a crucial task:
how to efficiently achieve entanglement of two SOQs? To this end we
refer to our previous investigations to create inter-spin entanglement
through SOC incorporating with Coulomb interaction \cite{Nan Zhao}.
In quantum electrodynamics (QED), the static Coulomb interaction is
a consequence of virtual photon exchange between two electric charges,
thus can be regraded as an external electric field exerted by one
electric charge on another. This implies that the SOQs can also feasibly
respond to the Coulomb interaction and thus cause a correlated motion
of two SOQs for generating quantum entanglement between SOQs. In ref.
\cite{Nan Zhao}, a scheme was proposed to create entanglement between
two local spins (rather than the SOQs) in two 2D quantum dots mediated
with SOC.

In this paper, we study an alternative scheme to generate entanglement
between two SOQs through Coulomb interaction. We will show that this
Coulomb interaction mediated entanglement can be optimized by the
strength of SOC: it is not the case the stronger the SOC is, higher
the entanglement is. More specifically, we consider a low dimensional
system made up by two electrons separately confined in two paralleled
nanowires and subjected to a weakly external magnetic field. By treating
the magnetic field as perturbation and modeling the Coulomb interaction
linearly nearby the equilibrium point, we derive an effective Hamiltonian
for two SOQs, with two qubits flip-flop induced by the Coulomb interaction.
It is found that the flip-flop rate as well as two qubits entanglement
(concurrence) changes non-monotonically with respect to SOC strength,
and they can be optimized by SOC. This discovery largely modifies
the intuitionistic observation that stronger coupling is more helpful
in creating highly entangled many body states.

This paper is organized as follow, in Sec. II we model the Coulomb
interaction as a linear coupling for two electrons trapped in two
quantum dots which are located in two parallel nanowires. In Sec.
III, we elucidate subspaces that can be used to encode SOQ under a
limiting situation where magnetic field can be treated as perturbation
to orbital motion as well as SOC. In Sec. IV, we discuss the generation
of the correlated flip-flop processes of two qubits as mediated by
the Coulomb interaction, and derive the effective Hamiltonian for
SOQs encoded using nearly degenerated ground states. Then we study
entanglement in the two SOQs system by calculating system dynamics
as well as time evolution of concurrence.

\section{MODEL SETUP FOR TWO PARALLEL NANOWIRES}

\begin{figure}
\begin{centering}
\includegraphics[scale=0.25]{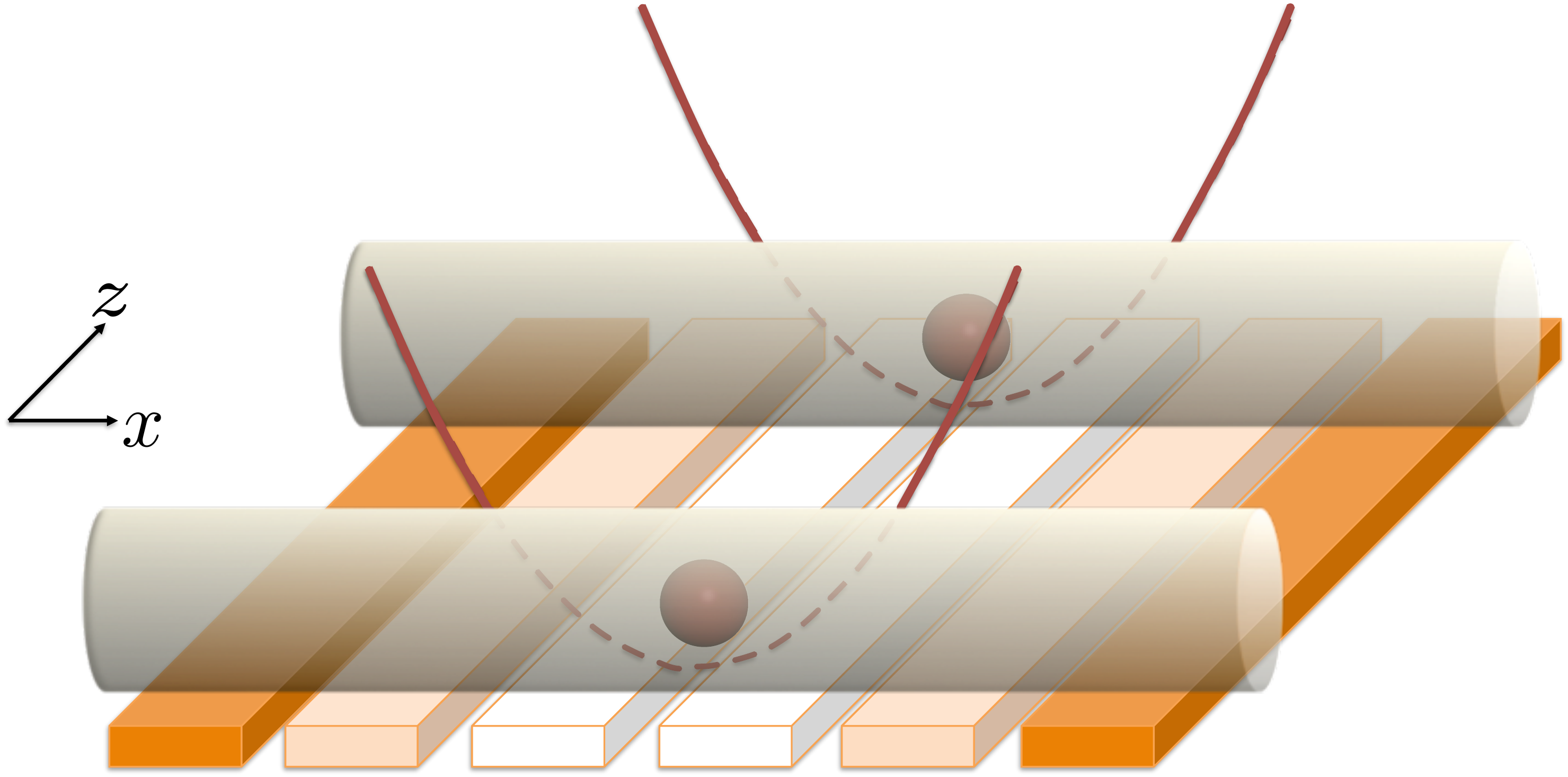}
\par\end{centering}

\caption{Schematic of two electrons confined in quantum dots which are located
separately in two 1D parallel nanowires. Electrodes beside nanowires
provide harmonic traps for electrons. Those trapping potentials are
represented by red parabolic curves.}
\end{figure}
We consider two parallel 1D nanowires placed along x-axis in the x-z
plane. Two electrons are confined by biased electrodes aside each
nanowire \cite{K Flensberg}, as schematically plotted in Fig. 1.
By treating the confining potentials as harmonic traps, we can write
down the Hamiltonian for the two electrons with SOCs 
\begin{equation}
H=\sum_{j=1,2}[\frac{p_{j}^{2}}{2m}+V_{j}+\alpha_{j}\sigma_{j}^{x}p_{j}+\mu_{j}\sigma_{j}^{z}]+V_{c},\label{1}
\end{equation}
where $j=1,2$ indicates confined electron in the front or at the
rear; $m$ is electron effective mass; $p_{j}$ and $x_{j}$ are momentum
and coordinate for j-th electron. We take the external trap to be
of harmonic oscillator, i.e., $V_{j}=m\bar{\omega}_{j}^{2}x_{j}^{2}/2$,
and $\bar{\omega}_{j}$ is characteristic oscillator frequency of
the trap; $\alpha_{j}$ is the Dresselhaus SOC strength and $2\mu_{j}$
is the Zeeman splitting for j-th electron \cite{Dresselhaus}. $V_{c}$
is Coulomb interaction between two electrons
\begin{equation}
V_{c}=\frac{e^{2}}{4\pi\varepsilon_{0}\varepsilon_{r}\sqrt{(x_{1}-x_{2})^{2}+(z_{1}-z_{2})^{2}}}.\label{2}
\end{equation}

Let $l_{j}=\sqrt{\hbar/m\bar{\omega}_{j}}$ be the characteristic
length of harmonic trap and $z_{0}$ be the separation between two
nanowires. In the case of strong confinement, i.e, $l_{j}\ll z_{0}$,$V_{c}$
can be further simplified \cite{Nan Zhao}. Specifically, we replace
$z_{1}-z_{2}$ in $V_{c}$ by $z_{0}$ and expand Eq.(\ref{2}) around
$x_{1}-x_{2}=0$. Then the potential is written up to second order
of $x_{1}-x_{2}$ as
\begin{equation}
V_{c}\approx V_{0}-\frac{1}{2}m\omega_{c}^{2}(x_{1}-x_{2})^{2},\label{3}
\end{equation}
where $V_{0}=e^{2}/(4\pi\varepsilon_{0}\varepsilon_{r}z_{0})$ and
$\omega_{c}=\sqrt{V_{0}/(mz_{0}^{2})}$.

With above approximation, the system reduces to two coupled harmonic
oscillators with SOC, with the Hamiltonian 
\begin{equation}
H=\sum_{j=1,2}[\frac{p_{j}^{2}}{2m}+\frac{1}{2}m\omega_{j}^{2}x_{j}^{2}+\alpha_{j}\sigma_{j}^{x}p_{j}+\mu_{j}\sigma_{j}^{z}]+m\omega_{c}^{2}x_{1}x_{2}.\label{4}
\end{equation}
Here, we have ignored the constant term $V_{0}$ and defined $\omega_{j}=\sqrt{\bar{\omega}_{j}^{2}-\omega_{c}^{2}}$.
We remark here that $\bar{\omega}_{j}>\omega_{c}$ always hold under
the pre-assumed condition $l_{j}\ll z_{0}$, thus $\omega_{j}$ is
a physically meaningful parameter.

In terms of the bosonic creation and annihilation operators $a_{j}^{\dagger}$
and $a_{j}$, $p_{j}$ and $x_{j}$ are rewritten as $p_{j}=i\sqrt{\hbar\omega_{j}m/2}(a_{j}^{\dagger}-a_{j})$
and $x_{j}=\sqrt{\hbar/(2m\omega_{j})}(a_{j}^{\dagger}+a_{j})$, respectively.
The Hamiltonian becomes $H=\sum_{j=1,2}(H_{j,0}+H_{j,1})+V$, where
\begin{equation}
H_{j,0}=\hbar\omega_{j}(a_{j}^{\dagger}a_{j}+\frac{1}{2})+i\xi_{j}(a_{j}^{\dagger}-a_{j})\sigma_{j}^{x}\label{8}
\end{equation}

\begin{equation}
V=g(a_{1}^{\dagger}+a_{1})(a_{2}^{\dagger}+a_{2}).\label{10}
\end{equation}
and $H_{j,1}=\mu_{j}\sigma_{j}^{z}$. In the above equations, $\xi_{j}=\sqrt{m\hbar\omega_{j}/2}\alpha_{j}$
are the rescaled SOC strength and we have introduced the coupling
constant $g=\hbar\omega_{c}^{2}/(2\sqrt{\omega_{1}\omega_{2}})$.

\section{DRESSED SPIN QUBIT IN MAGNETIC FIELD}

The existence of SOC in $H_{j,0}$ couples spatial motion of each
electron to its internal spin DOF. Therefore, it provides us with
an active method to manipulate spin states of electrons via controlling
over their spatial DOF \cite{SOQ control}. In this section, we study
the driven SOC system in the limit where $\mu_{j}\ll\hbar\omega_{j}$
and $\mu_{j}\ll\xi_{j}$, i.e., the strong SOC dominates over the
Zeeman effect.

First, let us analyze the energy level configuration of $H_{j,0}$.
The two fold degenerated eigenvectors of $H_{j,0}$ with the eigenvalue
$E_{j,n}=\hbar\omega_{j}(n-\eta_{j}^{2}+1/2)$ are $|n,\uparrow\rangle_{j}=|-i\eta_{j},n\rangle_{j}|\uparrow\rangle_{j}\mbox{ and }\mbox{ }|n,\downarrow\rangle_{j}=|i\eta_{j},n\rangle_{j}|\downarrow\rangle_{j}.$
Here, $|\alpha,n\rangle_{j}=D_{j}(\alpha)|n\rangle_{j}$ is displaced
Fock state created by displaced operator $D_{j}(\alpha)$ for $a_{j}$
and $a_{j}^{\dagger}$ \cite{D.F.Walls}; $\eta_{j}=\xi_{j}/(\hbar\omega_{j})$
is reduced SOC strength; $|\uparrow(\downarrow)\rangle_{j}$ are eigenstates
of $\sigma_{j}^{x}$ with eigenvalues $\pm1$. The following analysis
are identical for both electrons, thus we omit sub-index j for the
clarity of presentation in rest of this section.

\begin{figure}
\begin{centering}
\includegraphics[scale=0.28]{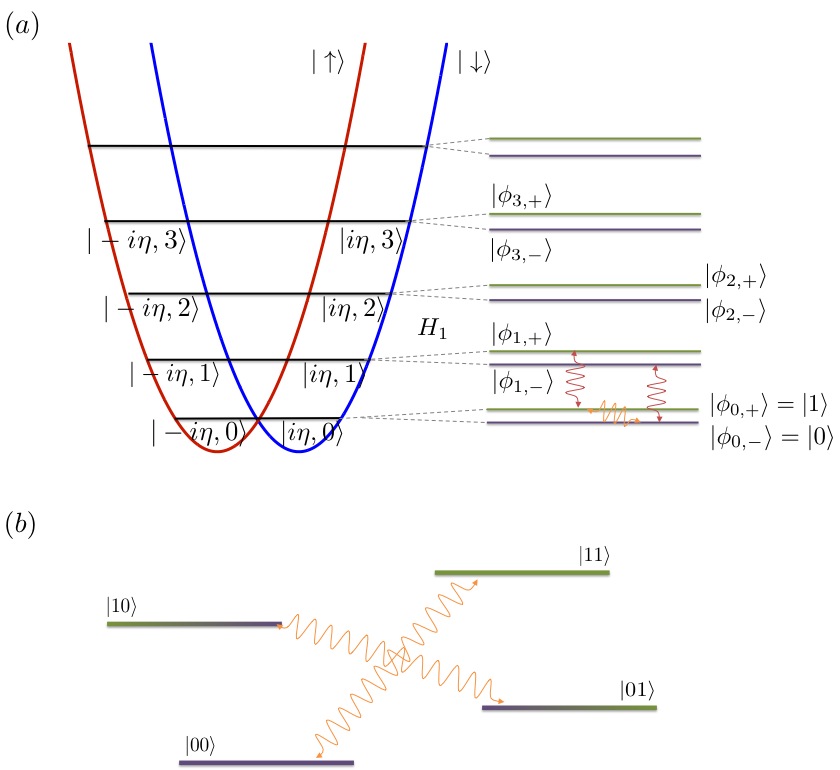}
\par\end{centering}

\caption{(a) Energy level structure of a spin with SOC in a weak magnetic field:
two parabolas indicate different harmonic traps felt by electron with
spin states $|\uparrow\rangle$ or $|\downarrow\rangle$. Black lines
inside parabolas denote eigenstates of $H_{0}$, which are the spin
dependent displaced Fock states. When $H_{1}$ is introduced, each
two degenerated levels split as shown on the right. A SOQ can be encoded
into the nearly degenerate ground states $|\phi_{0,\pm}\rangle$.
Transition between two qubit's basis can be achieved via external
a.c. electronic driving, which is shown by wavy arrows. The off-resonance
(red arrows) can be ignored under resonant driving. (b) Two SOQs coupled
by the linearized Coulomb interaction, the Coulomb interaction here
is of the same role as the a.c. field in the single qubit case. However,
only transitions between two pairs of basis are allowed due to generalized
parity selection rules.}
\end{figure}
As shown in Fig. 2a, since $H_{0}$ and $H_{1}$ do not commute with
each other, thus applying a magnetic field will then flip electron
spin and lift the previously degenerated levels. Generally the level
splitting will be proportional to the strength of the applied field
in weak field limit, so the two split levels are nearly degenerate
and consequently they can be used to encode a qubit, which is known
as SOQ.

In order to find out the level splitting as well as wave functions
for the dressed spin, we apply degenerate state perturbation theory
on $H_{1}$ by treating $\epsilon=\mu/(\hbar\omega)$ as a small parameter.
Following the procedure given in ref. \cite{Perturbation Ref}, we
first combine two degenerate eigenstates of $H_{0}$ such that $H_{1}$
is diagonal within their subspace. Since $\sigma^{z}$ flips $|\uparrow\rangle$
to $|\downarrow\rangle$, the diagonal matrix elements $\langle n,s|H_{1}|n,s\rangle$
vanish, where $s$ stands for $\uparrow$ or $\downarrow$, and off-diagonal
elements are
\begin{eqnarray}
\langle n,s|H_{1}|n,-s\rangle & = & \mu e^{-2\eta^{2}}L_{n}(4\eta^{2}).\label{19}
\end{eqnarray}
Here, $L_{n}(x)$ is Laguerre polynomials \cite{Laguerre poly}.

In the two dimensional subspace spanned by $\{|n,\uparrow\rangle,|n,\downarrow\rangle\}$,
$H_{1}$ is then written as
\begin{equation}
H_{1}=\mu e^{-2\eta^{2}}L_{n}(4\eta^{2})(|n,\uparrow\rangle\langle n,\downarrow|+h.c.),\label{21}
\end{equation}

The orthonormal eigenstates of above two-by-two matrix are $|n,\pm\rangle=\frac{1}{\sqrt{2}}(|n,\uparrow\rangle\pm|n,\downarrow\rangle)$,
the corresponding eigenvalues are
\begin{equation}
\delta E_{n,\pm}=\pm\mu e^{-2\eta^{2}}L_{n}(4\eta^{2}).\label{23}
\end{equation}
From the zero-th order wave function $|n,\pm\rangle$, we can use
the perturbation method to obtain the eigenvalue of $H_{0}+H_{1}$
as
\begin{equation}
E_{n,\pm}=E_{n}+\delta E_{n,\pm}
\end{equation}
where the first order correction $\delta E_{n,\pm}$ is given by Eq.(\ref{23}).
The corresponding first order eigenfunctions $|\phi_{n,\pm}^{(1)}\rangle$
are calculated in Appendix A. Two states with lowest energies $E_{0,\mp}$
are denoted by $|0\rangle\equiv|\phi_{0,-}^{(1)}\rangle$ and $|1\rangle\equiv|\phi_{0,+}^{(1)}\rangle$,
respectively. They are written explicitly as 
\begin{eqnarray}
|N\rangle & = & |0,(-1)^{N-1}\rangle+(-1)^{N}\sum_{m>0}[\lambda_{2m-1}|2m-1,(-1)^{N}\rangle\nonumber \\
 &  & +\lambda_{2m}|2m,(-1)^{N-1}\rangle],\mbox{ }(N=0,1)\label{Qubit basis}
\end{eqnarray}
where $\lambda_{m}=\epsilon\langle m|2i\eta\rangle/m$. 

Energy splitting between $|1\rangle$ and $|0\rangle$ is given by
$\Delta\equiv\delta E_{0,+}-\delta E_{0,-}=2\mu\exp(-2\eta^{2})\ll\hbar\omega$
since $\epsilon\ll1$. $\hbar\omega$ is level spacing between two
nearest unperturbed levels. If all relevant interactions are approximately
in resonant with $|0\rangle$ and $|1\rangle$, then the two states
can be used to encode quantum information as a SOQ. Aside from this,
it worths to remark another property of the SOQ. We notice that $|n,\pm\rangle$
looks similar to even and odd coherent states \cite{Even coherent state},
which process definite parities when the arguments of displacement
operator are purely imaginary. Here, as the direction of displacement
in $|n,\pm\rangle$ is entangled with spin orientation, they are eigenstates
of following generalized parity (GP) operator
\begin{equation}
\Lambda=\Pi\sigma^{z},\label{GPO}
\end{equation}
where $\Pi$ is the usual parity operator defined by $\Pi^{2}=1$
and $\Pi|x\rangle=|-x\rangle$ \cite{Cohen's QM book}. $p$, $x$
and $\sigma^{x}$ are odd under $\Lambda$, which means $\{\Lambda,p\}=0$
etc., and $\Lambda$ commutes with $H_{0}+H_{1}$. Therefore, SOQ
basis $|0\rangle$ and $|1\rangle$ can be simultaneously eigenstates
of $\Lambda$, i.e., $\Lambda|0\rangle=-|0\rangle$ and $\Lambda|1\rangle=|1\rangle$.
Furthermore, as shown in Appendix B, $|\phi_{n,+}^{(1)}\rangle$ and
$|\phi_{n+1,-}^{(1)}\rangle$ have the GPs that are different from
$|\phi_{n,-}^{(1)}\rangle$ and $|\phi_{n+1,+}^{(1)}\rangle$.

This property of $|\phi_{n,\pm}^{(1)}\rangle$ then indicates following
selection rules. First, for $H_{1}$, following transitions are forbidden
\begin{equation}
|\phi_{n,\pm}^{(1)}\rangle\leftrightarrow|\phi_{n+2m-1,\pm}^{(1)}\rangle,\mbox{ }|\phi_{n,\pm}^{(1)}\rangle\leftrightarrow|\phi_{n+2m,\mp}^{(1)}\rangle\label{29}
\end{equation}
since $H_{1}$ is even under $\Lambda$, i.e., $[H_{1},\Lambda]=0$.
Second, for the parity odd operator $a^{\dagger}+a\propto x$, following
transitions are forbidden
\begin{equation}
|\phi_{n,\pm}^{(1)}\rangle\leftrightarrow|\phi_{n+2m,\pm}^{(1)}\rangle,\mbox{ }|\phi_{n,\pm}^{(1)}\rangle\leftrightarrow|\phi_{n+2m-1,\mp}^{(1)}\rangle\label{selectrion rule for x}
\end{equation}

\section{ENTANGLEMENT BETWEEN TWO SOQS}

After elucidating the level structure and defining SOQ basis in previous
section, now we study how Coulomb interaction between two electrons
can be used to facilitate entanglement between SOQs. To this end,
we calculate the matrix elements of the linearized Coulomb interaction
Eq.(\ref{10}) in the basis of two SOQs subspace spanned by $\{|00\rangle,|01\rangle,|10\rangle,|11\rangle\}$,
where $|MN\rangle=|M\rangle_{1}\otimes|N\rangle_{2}$.

As shown in Appendix C, up to second order in $\epsilon_{j}$, matrix
elements of the Coulomb interaction are given by
\begin{eqnarray}
 &  & \langle N_{1}N_{2}|V|M_{1}M_{2}\rangle\nonumber \\
 & = & (-1)^{N_{1}+\bar{N}_{2}}\delta_{N_{1},\bar{M}_{1}}\delta_{N_{2},\bar{M}_{2}}J,\label{51}
\end{eqnarray}
where $\bar{N}=0$ (1) if $N=1$ (0), and the two qubits effective
flip-flop strength $J$ is obtained as
\begin{equation}
J=16g\epsilon_{1}\epsilon_{2}\eta_{1}\eta_{2}e^{-2(\eta_{1}^{2}+\eta_{2}^{2})}.\label{52}
\end{equation}
Existence of exponential decay term in $J$ indicates a non-monotone
behavior with respect to SOC. In fact, as shown in Fig. 3a, by setting
$\eta_{j}=1/\sqrt{2}$ the flip-flop strength is optimized to $J_{opt}=8g\epsilon_{1}\epsilon_{2}\exp(-2)$.

The possible transitions caused by the Coulomb interaction are illustrated
in Fig. 2b, where only transitions $|00\rangle\leftrightarrow|11\rangle$
and $|01\rangle\leftrightarrow|10\rangle$ are allowed. To illustrate
why others are forbidden, we recall a definition of the parity operator
as $\Pi_{j}=\exp(i\pi a_{j}^{\dagger}a_{j})$ \cite{definition of parity operator}.
Using this definition, $a_{j}$ and $a_{j}^{\dagger}$ can be shown
to be odd under the GP operation $\Lambda$. According to selection
rules similar to Eq.(\ref{selectrion rule for x}), GP of $|0\rangle$
and $|1\rangle$ must change after emitting or absorbing an phonon.
On the other hand, at first order the linearized Coulomb interaction
results in exchanging of an phonon between two electrons. Therefore,
the GP of each SOQ states must change as a result of the interaction.
Therefore, all transitions with only one SOQ changes its GP such as
$|00\rangle\leftrightarrow|01\rangle$ will vanish.

\begin{figure}
\begin{centering}
\includegraphics[scale=0.2]{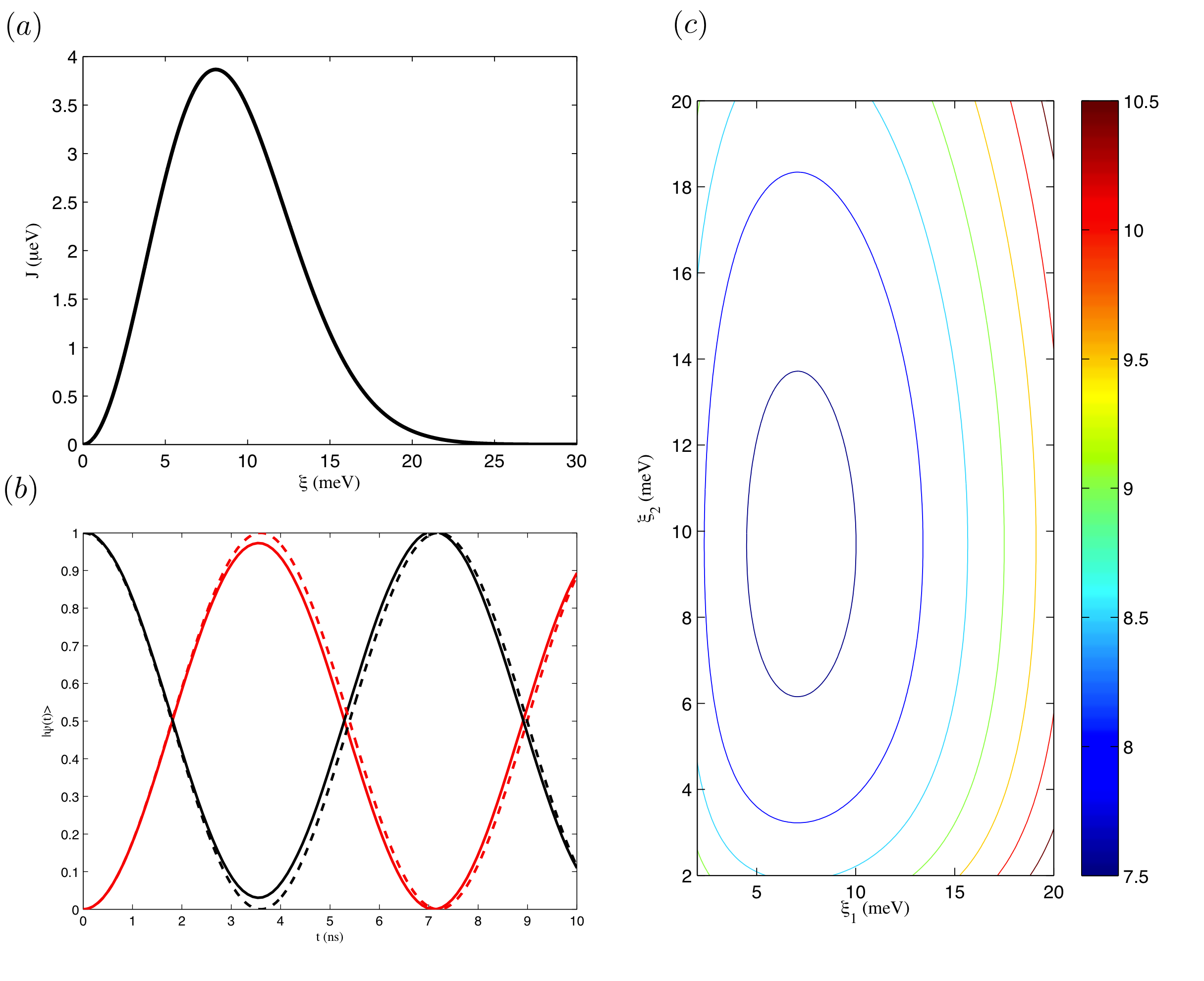}
\par\end{centering}

\caption{(a) Coulomb interaction induced SOQ effective flip-flop strength $J$
as a function of rescaled SOC strength $\xi_{1}=\xi_{2}=\xi$. (b)
Time evolution of two SOQs' wave function $|\psi(t)\rangle$ under
resonant condition $\Delta_{1}=\Delta_{2}$ and $|\psi(0)\rangle=|10\rangle$.
Red and black lines are $|\langle01|\psi(t)\rangle|^{2}$ and $|\langle10|\psi(t)\rangle|^{2}$,
respectively. Solid lines are calculated from Eq.(\ref{55}), while
dashed lines are exact result from $H$. (c) Minimal time $T$ needed
to complete a two qubits basis flip in resonant case as a function
of the rescaled SOC strength $\xi_{1}$ and $\xi_{2}$. The plot is
logarithm rescaled in z direction to stress the existence of minimal
point. Except for (c), parameters used in calculation are: $\bar{\omega}_{1}=20\mbox{meV}$,
$\bar{\omega}_{2}=15\mbox{meV}$, $\omega_{c}=5\mbox{meV}$, $\mu_{1}=\mu_{2}=1\mbox{meV}$
and $\eta_{1}=\eta_{2}=0.8$.}
\end{figure}
Introducing the following pseudo-spin operators $S_{j,+}=|1\rangle_{jj}\langle0|$,
$S_{j,-}=S_{j,+}^{\dagger}$ and $S_{j,z}=[S_{j,+},S_{j,-}]$, the
Hamiltonian of the two SOQ subsystem within the subspace of the two
qubits is written as
\begin{eqnarray}
H_{eff} & = & \sum_{j=1,2}\Delta_{j}S_{j,z}-J(S_{1,+}S_{2,+}-S_{1,+}S_{2,-}\nonumber \\
 &  & +h.c.).\label{55}
\end{eqnarray}
Here, we have ignored a constant energy shift $2(E_{1,0}+E_{2,0})$.

\begin{figure}
\begin{centering}
\includegraphics[scale=0.2]{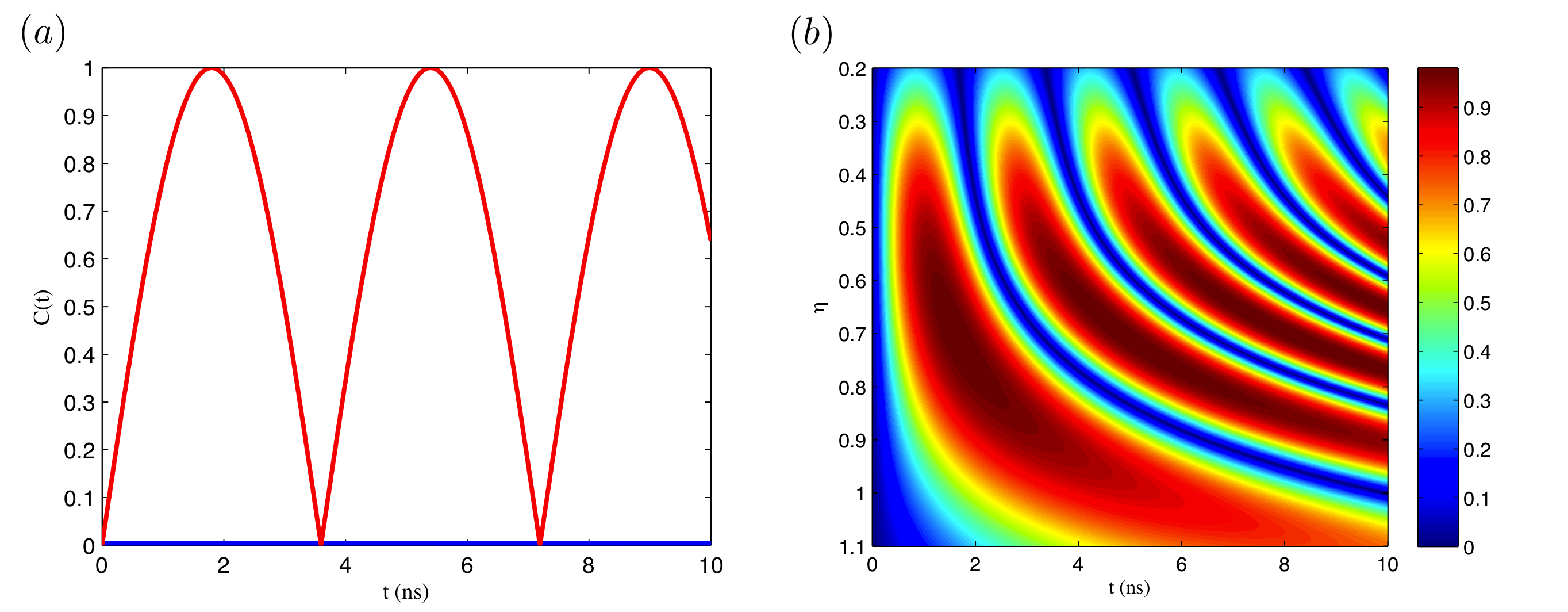}
\par\end{centering}

\caption{(a) Time evolution of concurrence $C(t)$ for two SOQs under initial
states $|10\rangle$ or $|01\rangle$ (red) and initial states $|00\rangle$
or $|11\rangle$ (blue) in resonant case, where maximum entangled
states can be generated at specific time. (b) Concurrence $C(t)$
for initial state $|10\rangle$ or $|01\rangle$ as functions of time
and reduced SOC strength $\eta_{1}=\eta_{2}=\eta$ for nearly resonant
case. Except $\mu_{2}=1.01\mbox{meV}$ for calculating (b), parameters
used in calculating those plots are the same as Fig. 3a,b.}
\end{figure}
With qubits flip-flop processes, entanglement between two SOQs can
be generated. To quantitatively discuss this, we study evolution dynamics
of the two qubits system as well as their concurrence. For the two
SOQs system initially in $|\psi(0)\rangle=|10\rangle$, the wave function
at time t is given by
\begin{equation}
|\psi(t)\rangle=(\cos\Omega t+i\sin\theta\sin\Omega t)|10\rangle+i\cos\theta\sin\Omega t|01\rangle
\end{equation}
Here, $\Omega=\sqrt{J^{2}+(\Delta_{1}-\Delta_{2})^{2}}$ and $\theta=\arctan[(\Delta_{1}-\Delta_{2})/J]$
are Rabi frequency and mixing angle, respectively.

Time evolution of $|\psi(t)\rangle$ is shown in Fig. 3b under resonant
condition $\Delta_{1}=\Delta_{2}$, which requires a homogenous magnetic
field over two nanowires. In this case $|10\rangle$ and $|01\rangle$
are degenerated in energy. Since $\theta=0$, thus the two degenerated
basis flip completely. Time needed for a complete flip $T=2\pi/J$
is plotted as function of $\xi_{j}$ and shown in Fig. 3c. Clearly
we see that by adjusting SOC, we can achieve the shortest operation
time.

We also calculated time evolution of concurrence $C(t)$ for two SOQs
to quantify degree of entanglement among them, using a formula for
two qubits in pure state \cite{Concurrence}, for initial state $|\psi(0)\rangle=|10\rangle$
or $|01\rangle$ 
\begin{equation}
C(t)=|\sin\Omega t|\sqrt{\sin^{2}2\theta\sin^{2}\Omega t+4\cos^{2}\theta\cos^{2}\Omega t}.\label{64}
\end{equation}

Notice that $C(t)=C(t+\pi/\Omega)$. When $\tan^{2}\theta\leq1$,
$C(t)$ always have two maximum values at 
\begin{equation}
t_{\pm}^{*}=\frac{1}{\Omega}\arccos[\pm\sqrt{\frac{1}{2}(1-\tan^{2}\theta)}]
\end{equation}
for $\Omega t\in[0,\pi]$. The corresponding maximum values are equal
and given by $C(t_{\pm}^{*})=1$. On the other hand, if $\tan^{2}\theta>1$,
then $C(t)$ has only one maximum at $t^{*}=\pi/(2\Omega)$ for  $\Omega t\in[0,\pi]$,
and the value is $C(t^{*})=|\sin2\theta|.$

While for the rest two initial states, i.e., $|11\rangle$ and $|00\rangle$,
the concurrence $C'(t)$ is also given by Eq.(\ref{64}) except $\Omega$
and $\theta$ being replaced by
\begin{equation}
\Omega'=\sqrt{J^{2}+(\Delta_{1}+\Delta_{2})^{2}},\mbox{ }\theta'=\arctan\frac{\Delta_{1}+\Delta_{2}}{J}.
\end{equation}

Time evolution of $C(t)$ are shown in Fig. 4a and Fig. 4b, where
we see that in resonant case the system can always evolve into a maximally
entangled state. For nearly resonant case, maximum value of $C(t)$
depends non-monotonically on SOC and can be optimized by tuning SOC.
However, it is hardly to observe any entanglement if initial state
is chosen as $|11\rangle$ or $|00\rangle$. Because $\Delta_{1}+\Delta_{2}\gg J$,
$\theta'$ approaches $\pi/2$ in this case , thus $C'(t)$ becomes
vanishingly small.

\section{CONCLUSION}

In this paper, we have studied entanglement induced by the Coulomb
interaction in a system of two electrons, which are separately trapped
in two 1D nanowires with SOC. We explicitly shown that how the presence
of magnetic field can enable the two electron spins to encode SOQs
in a regime where SOC and orbital motion of electrons dominate over
the Zeeman effect. The SOQ basis are shown to process the definite
GP, thus lead to selection rules under a large set of external driving
forces including the electronic field.

Based on the feasibly responding of SOQs to electric field \cite{KC Nowack},
we have shown that two qubits flip-flop can be effectively created
via the inter-electrons Coulomb interaction. In resonant case the
flip between two SOQs is perfect and the its period is found to depend
strongly on SOC and changing non-monotonically, which is in contrary
to an intuitive thinking. By studying the time evolution of concurrence,
we shown that entanglement among two SOQs in a nearly resonant case
can be optimized by adjusting the strength of SOC. 

Finally let us remark that it is possible to generate equally efficient
entanglement between $|00\rangle$ and $|11\rangle$ just by reversing
the direction of the external magnetic field on one SOQ setup, i.e.,
using inhomogenous magnetic fields. Since the two basis become energy-degenerated
if the fields on both SOQ setups are of the same magnitude. Meanwhile,
the coupling between all the rest pairs of basis can be achieved by
single qubit operations. 
\begin{acknowledgments}
This work is supported by National Natural Science Foundation of China
under Grants No.11121403, No.10935010, No.11074261, No. 91121015 and
and the National 973 program (Grant No. 2012CB922104 and No. 2014CB921402).
\end{acknowledgments}
\appendix

\section{SOQ wave functions by perturbation theory}

The first order eigenstates are calculated according to following
formula \cite{Perturbation Ref}
\begin{equation}
|\phi_{n,\pm}^{(1)}\rangle=|n,\pm\rangle+\sum_{m\ne n}\sum_{s'=\pm}\frac{\langle m,s'|H_{1}|n,\pm\rangle}{E_{n}-E_{m}}|m,s'\rangle,\label{A1}
\end{equation}
Matrix elements in the denominator can be rewritten as 
\begin{equation}
\langle m,\pm|H_{1}|n,\pm\rangle=\pm\mu\langle m|D(2i\eta)+D^{\dagger}(2i\eta)|n\rangle\label{24}
\end{equation}
as well as
\begin{equation}
\langle m,\mp|H_{1}|n,\pm\rangle=\pm\mu\langle m|D(2i\eta)-D^{\dagger}(2i\eta)|n\rangle.\label{25}
\end{equation}

Displacement operator under Fock states can be expressed in terms
of generalized Laguerre polynomials $L_{n}^{(m)}(x)$ \cite{Laguerre poly},
thus above matrix elements can be rewritten as 
\begin{eqnarray}
\langle m|D(2i\eta)\pm D^{\dagger}(2i\eta)|n\rangle & = & \frac{1}{\mu}(E_{n}-E_{m})\kappa_{m,n}^{(\pm)},\label{A4}
\end{eqnarray}
where $\kappa_{m,n}^{(\pm)}=\mu\zeta_{m,n}[1+(-1)^{|m-n|}]/(E_{n}-E_{m})$
and
\begin{equation}
\zeta_{m,n}=\frac{1}{2}(\frac{n!}{m!})^{f(m-n)}(2i\eta)^{|m-n|}L_{n}^{(|m-n|)}(4\eta^{2}),\label{A5}
\end{equation}
with auxiliary function $f(x)=1/2$ if $x\ge0$ and $-1/2$ otherwise. 

Insert Eqs.(\ref{A4}) and (\ref{A5}) back to Eq.(\ref{A1}), 
\begin{eqnarray}
 &  & |\phi_{n,\pm}^{(1)}\rangle\nonumber \\
 & = & |n,\pm\rangle\pm\sum_{m\ne n}(\kappa_{m,n}^{(+)}|m,\pm\rangle+\kappa_{m,n}^{(-)}|m,\mp\rangle).\label{A6}
\end{eqnarray}
For two lowest states where $n=0$, coefficients $\kappa_{m,0}^{(\pm)}$
are rewritten as
\begin{equation}
\kappa_{m,0}^{(\pm)}=-[1\pm(-1)^{m}]\lambda_{m},
\end{equation}
where $\lambda_{m}$ is defined in the main text. In this case, Eq.(\ref{A6})
recovers Eq.(\ref{Qubit basis}).

\section{GP of perturbed wave functions}

First notice that, for displaced Fock states 
\begin{eqnarray}
\Pi|\pm i\eta,2n\rangle & = & \int dx|-x\rangle\langle x|\pm i\eta,2n\rangle\nonumber \\
 & = & \int dxe^{\pm im\alpha x/\hbar}\varphi_{2n}(x)|-x\rangle\nonumber \\
 & = & \int dxe^{\mp im\alpha x/\hbar}\varphi_{2n}(x)|x\rangle\nonumber \\
 & = & |\mp i\eta,2n\rangle,
\end{eqnarray}
as well as
\begin{equation}
\Pi|\pm i\eta,2n-1\rangle=-|\mp i\eta,2n-1\rangle,
\end{equation}
where $\varphi_{n}(x)$ is eigenstates of 1D harmonic oscillator. 

When spin is included, following similar relations can be derived
\begin{eqnarray}
\Lambda|2n,\uparrow\rangle & = & \Lambda|-i\eta,2n\rangle|\uparrow\rangle\nonumber \\
 & = & |i\eta,2n\rangle|\downarrow\rangle\nonumber \\
 & = & |2n,\downarrow\rangle,
\end{eqnarray}
as well as
\begin{equation}
\begin{array}{c}
\Lambda|2n,\downarrow\rangle=|2n,\uparrow\rangle,\mbox{ }\Lambda|2n-1,\uparrow\rangle=-|2n-1,\downarrow\rangle\\
\Lambda|2n-1,\downarrow\rangle=-|2n-1,\uparrow\rangle
\end{array}\label{B11}
\end{equation}
Those relations then indicate that $|n,\pm\rangle$ are eigenstates
of the GP operator $\Lambda$
\begin{equation}
\begin{array}{c}
\Lambda|2n,+\rangle=|2n,+\rangle,\mbox{ }\Lambda|2n-1,-\rangle=|2n-1,-\rangle\\
\Lambda|2n,-\rangle=-|2n,-\rangle,\mbox{ }\Lambda|2n-1,+\rangle=-|2n-1,+\rangle
\end{array}\label{B12}
\end{equation}

From Appendix A, $\kappa_{m,n}^{(+)}$ ($\kappa_{m,n}^{(-)}$) is
non-zero only if $m$ and $n$ have same (opposite) oddness. Therefore,
Eq.(\ref{A6}) is rewritten as follow for $n$ even,
\begin{eqnarray}
 &  & |\phi_{2n,\pm}^{(1)}\rangle\nonumber \\
 & = & |2n,\pm\rangle\pm\sum_{m\ne n}(\kappa_{2m,2n}^{(+)}|2m,\pm\rangle+\kappa_{2m-1,2n}^{(-)}|2m-1,\mp\rangle),\nonumber \\
\end{eqnarray}
Together with Eq.(\ref{B12}), we then concluded that
\begin{equation}
\Lambda|\phi_{2n,+}^{(1)}\rangle=|\phi_{2n,+}^{(1)}\rangle,\mbox{ }\Lambda|\phi_{2n,-}^{(1)}\rangle=-|\phi_{2n,-}^{(1)}\rangle,
\end{equation}
and similarly analysis for the $n$ odd case gives
\begin{equation}
\Lambda|\phi_{2n-1,+}^{(1)}\rangle=-|\phi_{2n-1,+}^{(1)}\rangle,\mbox{ }\Lambda|\phi_{2n-1,-}^{(1)}\rangle=|\phi_{2n-1,-}^{(1)}\rangle.
\end{equation}

\section{Matrix elements of the Coulomb interaction}

In the subspace of two SOQs, matrix element of the linearized Coulomb
interaction is written as
\begin{eqnarray}
 &  & \langle N_{1}N_{2}|V|M_{1}M_{2}\rangle\nonumber \\
 & = & g\langle N_{1}|a_{1}^{\dagger}+a_{1}|M_{1}\rangle\langle N_{2}|a_{2}^{\dagger}+a_{2}|M_{2}\rangle.
\end{eqnarray}

Since $a_{j}^{\dagger}+a_{j}$ is odd under GP operation with respect
to j-th electron, thus $N_{j}$ and $M_{j}$ must be different in
order to have non-zero matrix elements (since $|0\rangle$ and $|1\rangle$
have different GP as shown in the main text). Therefore,
\begin{eqnarray}
 &  & \langle N_{1}N_{2}|V|M_{1}M_{2}\rangle\nonumber \\
 & = & \delta_{N_{1},\bar{M}_{1}}\delta_{N_{2},\bar{M}_{2}}J_{N_{1},N_{2}},\label{C17}
\end{eqnarray}
where $J_{mn}=g\langle m|a_{1}^{\dagger}+a_{1}|\bar{m}\rangle\langle n|a_{2}^{\dagger}+a_{2}|\bar{n}\rangle$.

Matrix element $\langle n|a^{\dagger}+a|\bar{n}\rangle$ can be calculated
directly from Eq.(\ref{Qubit basis}). Up to first order in $\epsilon$,
it is given by
\begin{equation}
\langle n|a^{\dagger}+a|\bar{n}\rangle=-2i(-1)^{n}\mbox{Im}\lambda_{1},
\end{equation}
Thus $J_{mn}$ can be rewritten as
\begin{eqnarray}
J_{mn} & = & 4(-1)^{m+n-1}g\prod_{j=1,2}\mbox{Im}\epsilon_{j}\langle1|2i\eta_{j}\rangle\nonumber \\
 & = & (-1)^{m+\bar{n}}16g\epsilon_{1}\epsilon_{2}\eta_{1}\eta_{2}e^{-2\eta_{1}^{2}-2\eta_{2}^{2}}.\label{C19}
\end{eqnarray}

Then Eq.(\ref{51}) can be recovered by inserting Eq.(\ref{C19})
back to Eq.(\ref{C17}). Notice that although $J_{mn}$ is second
order in $\epsilon_{j}$, actually we do not need to do perturbation
theory on wave function to that order. Because direct calculation
shows $\langle0,s|a^{\dagger}+a|0,s'\rangle=0$ even for $s\ne s'$,
then effect from the second order correction of wave function on matrix
element of the Coulomb interaction must be higher order and thus can
be neglected.


\begin{thebibliography}{10}
\bibitem{Entangle 1}Ryszard Horodecki, Pawe\l{} Horodecki, Micha\l{}
Horodecki, Karol Horodecki, Rev. Mod. Phys. 81, 865\textendash{}942
(2009).

\bibitem{SOQ 1}S. Nadj-Perge, S. M. Frolov, E. P. A. M. Bakkers,
and L. P. Kouwenhoven, Nature 468, 1084\textendash{}1087 (2010).

\bibitem{K Flensberg}Christian Flindt, Anders S. Sørensen, and Karsten
Flensberg, Phys. Rev. Lett. 97, 240501 (2006).

\bibitem{Rui Li}Rui Li, J. Q. You, C. P. Sun, and Franco Nori, arXiv:1304.3257.

\bibitem{EDSR}E. I. Rashba, and Al. L. Efros, Phys. Rev. Lett. 91,
126405 (2003).

\bibitem{Dresselhaus}G. Dresselhaus, Phys. Rev. 100, 580 (1955).

\bibitem{Nan Zhao}Nan Zhao, L. Zhong, Jia-Lin Zhu, and C. P. Sun,
Phys. Rev. B 74, 075307 (2006).

\bibitem{D.F.Walls}D.F. Walls and Gerard J. Milburn, \textit{Quantum
Optics} (Springer-Verlag, Berlin, Heidelberg, 1994).

\bibitem{SOQ control}Christian Flindt, Anders S Sørensen and Karsten
Flensberg, J. Phys.: Conf. Ser. 61, 302 (2007).

\bibitem{Perturbation Ref}J. J. Sakurai, Jim Napolitano, \textit{Modern
Quantum Mechanics 2nd ed}., (Addison-Wesley, 2011).

\bibitem{Laguerre poly} A. Ferraro, S. Olivares, M. G. Paris, \textit{Gaussian
States in Quantum Information}, Napoli. Series on physics and astrophys.
(2005).

\bibitem{Concurrence}Charles H. Bennett, David P. DiVincenzo, John
A. Smolin, and William K. Wootters, Phys. Rev. A 54, 3824 (1996).

\bibitem{Even coherent state}V.V. Dodonov, I.A. Malkin, V.I. Man'ko,
Physica (Utrecht) 72, 597 (1974).

\bibitem{KC Nowack}K. C. Nowack, F. H. L. Koppens, Yu. V. Nazarov,
L. M. K. Vandersypen, Science 318, 1430 (2007).

\bibitem{Cohen's QM book}Claude Cohen-Tannoudji, Bernard Diu, Franck
Laloë, \textit{Quantum mechanics}, (Wiley-VCH, 1992).

\bibitem{definition of parity operator}Antonino Messina, and Gheorghe
Draganescu, ArXiv: 1306. 2524.\end{thebibliography}
\end{document}